# Algorithm for Achieving Consensus Over Conflicting Rumors: Convergence Analysis and Applications

Ismail Elouafiq, *Student Member, IEEE*, Amine Semma, *Student Member, IEEE*,

*Abstract*—Motivated by the large expansion in the study of social networks, this paper deals with the problem of multiple messages spreading over the same network using gossip algorithms. Given two messages distributed over some nodes of the graph, we first investigate the final distribution of the messages given an initial state. Then, an algorithm is presented to achieve consensus over one of the messages. Finally, a game theoretical application and an analogy with word-of-mouth marketing are outlined.

## I. Introduction

In recent years, large decentralized distributed systems such as sensor and wireless networks require the design of communication schemes that satisfy scalability, robustness and graceful degradation. Consequently, information dissemination and message spreading algorithms have generated a huge interest.

The problem of broadcasting may be defined as spreading some information over a graph that is unknown to its nodes. Gossip algorithms are a set of algorithms that are used to solve such problems. *Randomized gossip* is one of the most widely used forms of gossip that have gained prominence for the simplicity of its protocol. In a random gossip based algorithm, nodes repeatedly call a random neighbor to transfer their messages. At the end of the process, the information should be spread across the network structure for the algorithm to converge.

In this paper, based on gossip algorithms we investigate the spreading of rumors over a social network. We will start by describing the case of two conflicting messages spreading by considering an algorithm that has already been proposed in literature [1]. Given a state of the graph where the two messages are held by some of the nodes, we analyze the final state of the graph based on a deterministic model based on the expectation of a Markov Chain. After the two messages are spread over the network a consensus could be achieved over one of the two messages. For that reason, a simple and efficient consensus-based algorithm is proposed to attain this goal.

One other important aspect of gossip algorithms is their ability to model human behavior. For instance, in a social network where the agents are assumed to make rational decisions these agents base their belief choices on likelihood criteria. For this reason, we decided to investigate the use of the consensus-based gossip algorithm in game theory, more precisely in a voting scenario where multiple agents act according to their local knowledge. Based on that scenario, we propose an application of gossip algorithms in modeling the Word of Mouth (WoM) marketing strategy where customers contribute in the marketing strategy of a given product or service.

The rest of this paper is organized as follows. In Section II, a protocol modeling the case of two messages being forwarded over the same network is proposed and analyzed. Then, an algorithm for achieving consensus over two spread messages is described in Section III. Section IV provides two different applications of gossip algorithms, one in game theory and the other in marketing . Finally, we finish with some conclusions and future directions in Section VII.

## II. Two Conflicting Rumors Spreading over the Same Network

We investigate the problem of spreading two conflicting messages *m1* and *m2* on a social network.

For concreteness, let us begin with a review of a message spreading algorithm. In a social network, represented by a graph, informed nodes aim at sending their message to their neighbors. We propose an asynchronous gossip based model. At each round, a node $i$ is chosen at random over the network. If $i$ is holding one of the two messages, i selects a neighbor $j$ uniformly at random and informs it. We point out that this approach does not give a realistic representation of the problem since a node may lose interest in spreading the message if it tries to inform another node already knowing it. To tackle this issue, we propose to study a different model proposed in [1].

*A. Algorithm description :*

Let $G=(V,E)$ be a complete graph where $|V|=n$ and $E=V \times V$ and let $l$ be a positive integer $l \in \mathbb{N}$. The algorithm proceeds as follows :

In the beginning, two subsets of $V$, $I_1$ and $I_2$, are respectively informed by *m1* and *m2*. At each round :

- A node $i$ is chosen at random over $V$. if $i$ holds a message :
  - $i$ chooses uniformly at random a neighbor $j$ with a probability $1/(n-1)$.
  - If node $j$ doesn't have the message (*susceptible node*), node $i$ informs it ($j$ becomes an m1 *infective node*). Else, $i$ increments a counter $C_i$ with one ($C_i$ counts the unnecessary calls made by $i$, their initial value is 0).
  - If $C_i = l$, node $i$ stops spreading the message and becomes a removed node.

To start, we will assume that l=1. According to the algorithm description, nodes can be in one of the five following states:

- *State* 1: Has (*m1*) and is spreading it "*m1 infective*".

- *State* 2: Has (*m2*) and is spreading it "*m2 infective*".
- *State* 3: Has (*m1*) and stopped spreading it " *m1 removed*".
- *State* 4: Has (*m2*) and stopped spreading it "*m2 removed*".
- *State* 5: Doesn't have a message "*susceptible*"

Following these five states, at each step *k* let *I1(k), I2(k), S(k), R1(k)* and *R2(k)* denotes number of nodes in state 1, 2, 3, 4 and 5 respectively.

*B. Markov Chain Model*

If we denote by *X(k)* the vector *(I1(k), I2(k), S(k), R1(k), R2(k))$^T$*, the communication step depends only on *X(k)* the state of the nodes at *k*, thus *X(k+1)* depends only on *X(k)*. The process can then be modeled with a DTMC (Discrete Time Markov Chain).

Let us compute its transitions probabilities. Actually, given a state S(k), there are 5 possible transitions :

- $p_0(k)=Pr(X(k)\rightarrow(I(k)+1, I2(k), Y(k)-1, R1(k), R2(k)))$
$$p_0(k) = \frac{S(k)+R1(k)+R2(k)}{N} \quad (1)$$
- $p^1_+(k)=Pr(X(k)\rightarrow(I1(k)+1, I2(k), Y(k)-1,R1(k),R2(k)))$
$$p^1_{+(k)} = \frac{I1(k).S(k)}{N(N-1)} \quad (2)$$
- $p^1_-(k)=Pr(X(k)\rightarrow(I1(k)-1, I2(k), Y(k), R1(k)+1,R2(k)))$
$$p^1_{+(k)} = \frac{I1(k).(N-S(k))}{N(N-1)} \quad (3)$$
- $p^2_+(k)=Pr(X(k)\rightarrow(I(k), I2(k)+1, Y(k)-1, R1(k), R2(k)))$
$$p^2_{+(k)} = \frac{I2(k).S(k)}{N(N-1)} \quad (4)$$
- $p^2_-(k)=Pr(X(k)\rightarrow(I(k), I2(k)-1, Y(k), R1(k), R2(k)+1))$
$$p^2_{+(k)} = \frac{I2(k).(N-S(k))}{N(N-1)} \quad (5)$$

*C. Deterministic model*

The *DTMC* introduced above is reducible and transient as states of the form *(0, 0, j, r1)* are absorbing states. In order to have a steady-state distribution, the first *DTMC* is modified as follows: Given an initial state *(n1, n2, n-n1-n2, 0)*, and for any *j* and *r1 >= n1, p0, 0, j, r1(n1, n2, N-n1-n2, 0) = 1*.

However, the computation of its steady-state has a high complexity and so we will compute its conditional expectation and then deduce a deterministic model.

The expectation calculi give :

$$E[I1(k+1)|I1(k)] = I1(k) + \frac{I1(k).(2.S(k)-N)}{N(N-1)} \quad (6)$$

$$E[I2(k+1)|I2(k)] = I2(k) + \frac{I2(k).(2.S(k)-N)}{N(N-1)} \quad (7)$$

$$E[S(k+1)|S(k)] = S(k) - \frac{S(k).(I1(k)+I2(k))}{N(N-1)} \quad (8)$$

$$E[R1(k+1)|R1(k)] = R1(k) + \frac{I1(k).(N-S(k))}{N(N-1)} \quad (9)$$

$$E[R2(k+1)|R2(k)] = R2(k) + \frac{I2(k).(N-S(k))}{N(N-1)} \quad (10)$$

Thus, if we denote by *i1, i2, s, r1* and *r2* respectively "*infective m1*", "*infective m2*", "*susceptible*", "*removed m1*" and "*removed m2*" nodes, and using (6), (7), (8), (9), (10) and the two assumptions:
- *N~N-1* for high *N*
- Time intervals between the Poisson clock ticks is neglected (*i.e.: k=t*).

A deterministic model is deduced :

$$\frac{di1(t)}{dt} = i1(t)(2s(t)-1) \quad (11)$$

$$\frac{di2(t)}{dt} = i2(t)(2s(t)-1) \quad (12)$$

$$\frac{ds(t)}{dt} = -s(t)(i1(t)+i2(t)) \quad (13)$$

$$\frac{dr1(t)}{dt} = i1(t)(1-s(t)) \quad (14)$$

$$\frac{dr2(t)}{dt} = i2(t)(1-s(t)) \quad (15)$$

Furthermore, we can note that :

$$\frac{di1(t)}{di2(t)} = \frac{i1(t)}{i2(t)} \quad \text{and} \quad \frac{dr1(t)}{dr2(t)} = \frac{i1(t)}{i2(t)} \quad (16)$$

which gives an interesting relation between *i1(t)* and *i2(t), r1(t)* and *r2(t)* :

$$i1(t) = \frac{i1(0)}{i1(0)+i2(0)}.i(t) \quad (17)$$

$$r1(t) = \frac{i1(0)}{i1(0)+i2(0)}.r(t) \quad (18)$$

where *i(t)=i1(t)+i2(t)* and *r(t)=r1(t)+r2(t)*.

As a consequence, the problem is reduced to one message spreading over a social network. Then, we reduce (11), (12), (13), (14) and (15) to two equations. Moreover, in [3], it is showed that even if *l>1* equations still hold by adding a multiplying coefficient *1/l*. Let *s, i* and *r* denote respectively the fractions of susceptible, infective, and removed individuals, such that *s + i + r = 1* :

$$\frac{ds(t)}{dt} = -si \quad (19)$$

$$\frac{di(t)}{dt} = si + \frac{1}{l}(1-s)i \quad (20)$$

The solution of this couple of equations is [3]:

$$i(s) = \frac{l+1}{l}(1-s) + \frac{1}{l}\log(s) \quad (21)$$

This gives us a first indication on how *i* changes with *s* (Figure 1).

The algorithm stops when all the infective nodes stop spreading the message, *i.e: i = 0*. According to equation (21), *i(s)* is zero when :

$$s = \exp(-(l+1)(1-s)) \quad (22)$$

Which gives an implicit solution of the equation. Here are some theoretical results for the reach (number of nodes that end by having the message) given two different values of k :
- For *l=1, s=20%*.
- For *l=2, s=6%*.

When in the final state, we can deduce easily *r1* and *r2* by multiplying *r* by the initial coefficient respectively i1*(0)/i(0)* and *i2(0)/i(0)*.

### D. Simulation results:

The first simulation (Figure 1) shows the evolution of *i* as a function of *s* for different *l* values. We can note that as the value of *l* increases the curve gets closer to a linear shape. This is justified by the fact that (21) becomes *i=1-s* which is in fact a linear function. Moreover, the plot of the theoretical curve fits the simulation results when *l=1* which validates the deterministic model.

Then, using the Monte Carlo method, the simulation results in Figure 2 were obtained. Figure 2 shows the mean of the difference between the two messages sets cardinalities as a function of the initial difference. We can note the linear shape of the curve as expected with the deterministic model in the last section.

### E. Multiple messages

In the case of multiple messages, we assume that we are concerned by only one of these messages and the rest are considered as disrupting messages. Hence, the problem can be reduced to a two conflicting messages spreading over the same network (one for the valid message and the other englobing all the adversary messages). All the results presented in this section could be applied.

## III. CONSENSUS OVER TWO DIFFUSED MESSAGES

We assume that the initial state of the graph is as follows :
- *n1* nodes received message (*g1*).
- *n2* nodes received message (*g2*).
- All the nodes received a message : *n1 + n2 = n*.

Then, we want all the nodes to agree on one of the messages. We start by considering an asynchronous algorithm that uses two increasing counters and we show that it is equivalent to a seemingly simpler algorithm that only uses one counter averaged at each step.

First, we assume that each node *i* holds two counters for received messages (a counter $C_i$ for the messages corresponding to his own message, and a second counter $C'_i$ for those corresponding to the other message).

All the counters start by a value of zero. At each step *k*, a randomly chosen node *i* is woken up. Then, a neighbor *j* is chosen uniformly at random. Both nodes exchange messages and counters, then update their counters as follows :
- if the received message is different from the held message, its counter is incremented by the value of the received counter ($C'_i(k+1)=C'_j(k+1)=C'_i(k)+C_j(k)$). The new message is stored.

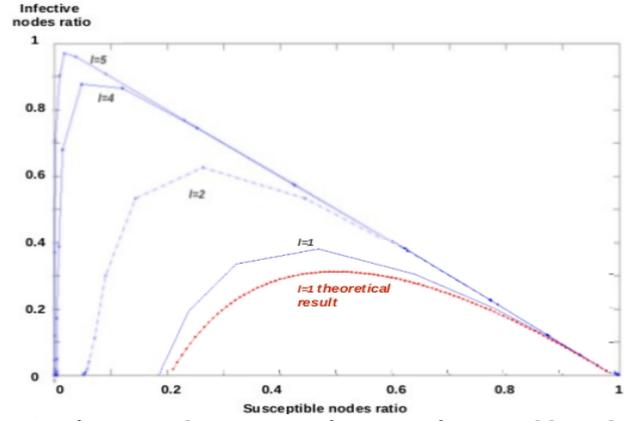

**Fig. 1** Infective nodes ratio as a function of susceptible nodes ratio (Theoretical result for l= 1 and simulation results for l=1,2,4,5)

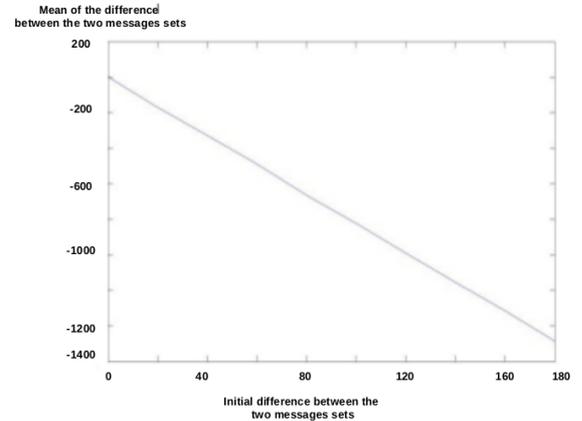

**Fig. 2** Mean of the difference between the two messages sets as a function of the initial difference : comparison between simulation and theoretical results (n=5000 and n1+n2=200).

- if the received message corresponds to the held message, the counter corresponding to the held message is incremented by the value of the received counter ($C_j(k+1)=C_i(k+1)=C_i(k)+C_j(k)$).

For a node *i* at step *k*, to choose the most relevant message, *i* simply compares his two counters. Comparing the counters is equivalent to subtracting the value of $C'_i$ from that of $C_i$.

Thus we propose the following algorithm to study. We give to each node *i* a counter $C_i$ and proceed by averaging the counters to achieve consensus as described below.

### A. Algorithm description :

Initialization step :
for *i* in {*1,..,n*}

$$C_i = \begin{cases} 1, & \text{if } i\,holds\,(g1) \\ -1, & \text{if } i\,holds\,(g2) \end{cases}$$

> Communication step :
> At the rate of a Poisson process (step $k$) , for each node $i$ :
> - $i$ wakes up.
> - $i$ chooses a neighbor $j$ uniformly at random.
> - Both nodes update their counters :
>   $C_j(k+1)=C_i(k+1)=(1/2)( C_i(k)+C_j(k))$
> - Both $i$ and $j$ sleep.

## B. Convergence analysis :

In this section, we will study the convergence of the randomized gossip algorithm. In the algorithm described above nodes proceed by updating their local counter at each step. Let $N_i$ be the set of $i$'s neighbors, $n_i=|N_i|$ the number of $i$'s neighbors and $C(k)$ the vector for which entries are the counters $C_i(k)$ at each time slot $k$. Thus, the update can be modeled linearly by the following equation:

$$C(k+1)=W(k).C(k) \quad (23)$$

where, when node $i$ chooses node $j$ from $N_i$:

$$W(k)=I-\frac{(e_i-e_j)(e_i-e_j)^T}{2} \quad (24)$$

with probability $\frac{1}{n.n_i}$, where $e_i$ is an $n \times 1$ unit vector with the $i^{th}$ component equal to one.

$W(k)$ must satisfy some constraints according to [2]. These constraints are imposed by the gossip algorithm and the graph topology.

If nodes $i$ and $j$ are not neighbors, $W_{ij}(k)$ must be zero. Further, since every node can communicate with only one of its neighbors per time slot, each column of $W(k)$ can have only one non-zero entry other than the diagonal entry.

In each iteration, the averaging computation impose the preservation of the sums : this means that $\mathbf{1}^T.W(k) = \mathbf{1}^T$, where $\mathbf{1}$ denotes the vector of all ones. Also, the vector of averages must be a fixed point of the iteration, i.e. $W(k).\mathbf{1} = \mathbf{1}$.

Since the choice of $i$ and $j$ are independent of the time slot, $W(k)$ matrices are *IID*. Secondly, from (23), we can find by iteration that :

$$C(k)=(\prod_{l=0}^{l=k-1} W(l)).C(0)=P(k-1).C(0) \quad (25)$$

Hence, if $C(k)$ must converge to $C_{ave}.\mathbf{1}$ we must have :

$$\lim_{k\to\infty} P(k)=\frac{\mathbf{1.1}^T}{n} \quad (26)$$

In the next sections, the convergence to the initial counters mean will be proven.

*1) Convergence in expectation:*
Let $W = E[W(k)]$. Under the following conditions :
  (a1)- $\mathbf{1}^T.W = \mathbf{1}^T$
  (a2)- $W.\mathbf{1} = \mathbf{1}$
  (a3)- $\rho(W-\frac{\mathbf{1.1}^T}{n})<1$ where $\rho(.)$ is the spectral radius of a matrix.

We have : $\lim_{k\to\infty} E(P(k))=\frac{\mathbf{1.1}^T}{n} \quad (27)$

Then: $\lim_{k\to\infty} E(P(k))=\frac{\sum_{i=1}^{i=n} C_i(0)}{n} \quad (28)$

*2) Convergence of the second moment :*
The convergence of the second moment is investigated in this section to quantify the convergence rate of $C(k)$ to $C_{ave}$.

To obtain it, lets consider the error : $N(k)=C(k) - C_{ave}.\mathbf{1}$. Considering the evolution of $N(k)$, we can easily demonstrate that :

$$N(k+1)=W(k).N(k) \quad (29)$$

So, $N(k)$ evolves with the same linear system as $C(t)$. Hence we can write :

$$E[N(k+1)^T.N(k+1)|N(k)] = N(k)^T.E[W(k)^T.W(k)].N(k) \quad (30)$$

Using the fact that $W(k)$ is doubly stochastic and so is $W$, and the orthogonality between $N(t)$ and $\mathbf{1}$, we can demonstrate that :

$$E[N(k)^T N(k)]\leq \lambda_2^{2t}(E[W^{TW}]) N(0) \quad (31)$$

Hence, since $\lambda_2<1$ (second largest *eigen value* of $W$), the expectation of the error converge to zero when $k$ approaches *infinity*.

*3) High probability bounds on averaging time :*
In [2], an upper and lower bounds are demonstrated.

**Theorem 1 :[2]** Having a gossip algorithms with an initial state $C(0)$:

For $k \geq 6.K^*(\varepsilon)$ : $Pr(\frac{\|C(k)-C_{ave}\|}{\|C(0)\|}\geq \epsilon)\leq \epsilon \quad (32)$

For $k < K^*(\varepsilon)$ : $Pr(\frac{\|C(k)-C_{ave}\|}{\|C(0)\|}\geq \epsilon)> \epsilon \quad (33)$

Where : $K^*(\varepsilon)=\frac{\log(\epsilon^{-1})}{2.\log(\lambda_2(W)^{-1})} \quad (34)$

These are results for averaging consensus. However, the consensus studied her is to reach all the nodes to have the same decision on the message they spread. It's clearly a consensus as it's a fixed point of the algorithm.

To reach that, a sufficient but not necessary condition can be implemented as a stopping criterion :

$$\|C(k)-C_{ave}\|<C_{ave} \quad (35)$$

This means that :

$$\forall 1\leq i \leq n, \quad |C_i(k)-C_{ave}|<C_{ave} \quad (36)$$

Hence, with (4), we can obtain the corresponding $\varepsilon$ for the convergence. For $\epsilon=\frac{\|C_{ave}\|}{\|C(o)\|}=\frac{|\sum_{i=0}^{n} C_i(0)|}{n\sqrt{(n)}}=\frac{|n_1-n_2|}{n\sqrt{(n)}} \quad (37)$

Note that $\varepsilon$ is an increasing linear function of the initial difference between the number of *g1* and *g2* holders. We have:

$$Pr(C_i(k) \text{ have the same sign})\geq 1-\frac{|n_1-n_2|}{n\sqrt{(n)}} \quad (38)$$

for every $k \geq \dfrac{3.\log(\frac{\|C(o)\|}{\|C_{ave}\|})}{\log(\lambda_2(W)^{-1})}$ (39)

and :

$$Pr(\text{All } C_i(k) \text{ have the same sign}) < 1 - \dfrac{|n_1 - n_2|}{n\sqrt{(n)}} \quad (40)$$

for every $k < \dfrac{\log(\frac{\|C(o)\|}{\|C_{ave}\|})}{2.\log(\lambda_2(W)^{-1})}$ (41)

Which gives the consensus reach bounds of the algorithm.

*C. Simulation results*

First, we simulate (Figure 3) the evolution of the number of nodes holding each one of the two messages (*g1*) and (*g2*). The two simulations concern a complete graph with 1000 nodes. This simulation shows that the algorithm converges and the nodes end up agreeing on the most relevant message (*i.e*: the one with the higher initial number of holders).

Secondly, the curves in Figure 4 show the evolution of the distance $\|C(k) - C_{ave}\|$. Foremost, we can see that the convergence is reached even before getting a distance of zero between the counter's vector and $C_{ave}$, which can be justified by (26). In fact, the convergence is a convergence of the signs of the counters rather than the counters themselves. Then, it is also clear ( Figure 4) that when the initial settings of *n1* and *n2* are farther from each other the convergence rate is higher, which matches the theoretical results.

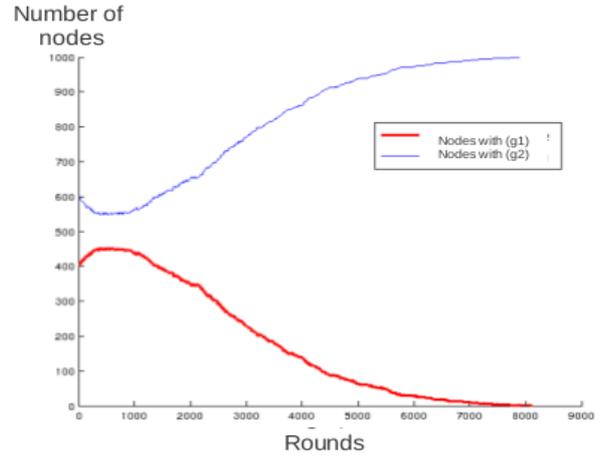

**Fig. 3** Evolution of holders of each message (g1: lower curve, g2: upper curve) at each round (initial values: n1=400 and n2=600)

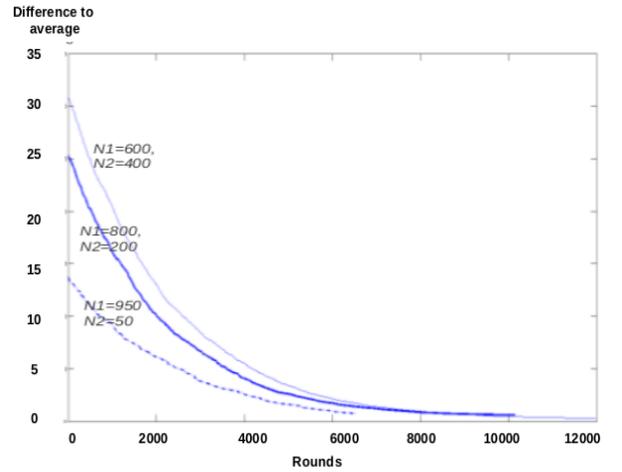

**Fig. 4** Distance to the expected average at each round k(i.e.: $\|C(k) - C_{ave}\|$ ). Three initial settings are distinguished

## IV. APPLICATION

*A. A repeated game of distributed voting:*

We consider a repeated game where players can vote for one of two possible candidates. Each agent prefers one of the two candidates over the other. However, a player is better off voting for the candidate that is going to win anyway. Hence, we define $u_k(i)$, the payoff of each player *i* at a step k, as follows:

$$u_k(i) = (\text{Sum of all the agents agreeing with } i)$$

Thus, the set of agents *N* is partitioned so that we have two subsets $N_1$ and $N_2$ containing respectively the agents agreeing over the choice $\chi_1$ or $\chi_2$.

Since the agents are unaware of the choices of other agents they cannot know what their payoffs are at a certain step. Moreover, we assume that at each step, to agents are able to contact each other. The goal is to find a strategy that will help all the agents maximize their payoffs.

Proceeding as shown in the algorithm provided in Section III. All the players hold a counter that is initialized at the beginning of the game.

At a step *k*, we assume that no agent is able to observe his payoff. Instead, each couple of agents *i* and *j* who came in contact with each other can observe the state of their counters.

According to the results of the previous section, consensus over the choice of a candidate is reached with high probability.

*B. Application to word-of-mouth marketing*

Direct marketing deals with separate events: each email or advertisement is considered as a separate deal. But in a community where people are related to each other, all of these notions are connected.

We analyze the case where two conflicting products are spreading in a community. There aren't many distinct thresholds for spreading two products in a community: one of the products will win at the very end. But in general, the spreading of the products stops before this state is achieved.

The individuals proceed as shown in the algorithm described in Section III. We start by considering that all the nodes (individuals) are connected to each other. If the graph is not complete, the counters will be weighted proportionally to the centrality of the nodes. We can choose *betweenness* as a measure of centrality. The intuition behind this approach is that a node by which transits more information is more likely to have an impact on the other nodes. The only thing that changes is the initialization step and we are brought back to the complete graph structure. Thus, we consider a set of agents $V$ connected in a complete undirected graph $G = (V, E)$.

Instead of initializing all the nodes at the same value of the counter, a higher value should be assigned to the counters of the nodes that are more convincing.

First, we assume that more or less convincing nodes are distributed according to a normal distribution. In other words, in the initialization step: the random variable that assigns a value to a counter follows a Gaussian distribution.

According to Section III, at each step, each counter is a linear combination of the initial values of the counters. Consequently, at each step, the vector $C(k)$ follows a normal distribution. Moreover, the values of the counters converge to the mean of the initial distribution. Hence, the mean of the final distribution is equal to the initial mean. However, the closer this initial mean is to zero, the slower is the convergence of the algorithm.

**Simulation results:**

Experimental results (figure 5) show that the final distribution is as expected a normal distribution.

Changing the initial variance of the distribution only impacts the convergence rate.

We simulate a graph where the distribution has a mean that has a negative value close to zero. As shown in Figure 5, the individuals end up having the same preference, *i.e.:* all the counters are negative.

However, in the real world the spreading process stops before convergence is reached since new products come out.

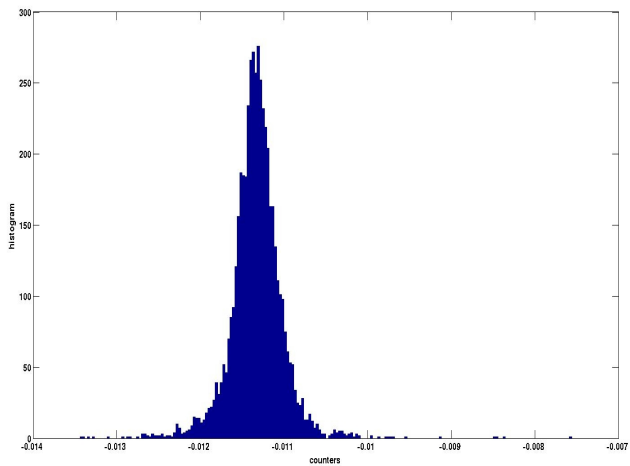

**Fig. 5:** Last distribution of the counters (the values are condensed around a value close to -0.0112)

## V. CONCLUSION AND FUTURE WORK

Gossip algorithms are an efficient tool to model rumor spreading over network structures. The results presented in this paper give an overview of the way rumor forwarding proceeds in a social network.

In the case of multiple conflicting messages, our main result is that the number of holders of a given message evolves almost proportionally to the sum of all the messages holders with a coefficient that is equal to the initial number of holders.

Furthermore, the consensus based gossip algorithm makes it possible to lead the nodes towards choosing the same belief. Of course, this is does not represent realistic scenarios as many parameters were not considered. For instance, there might exist some non-cooperating agents among the nodes.

A game theoretical approach shows that agents can maximize their payoffs based only on their local knowledge and agree on the same choice in a voting game.

A word-of-mouth marketing model shows how, in a community where two conflicting ideas start spreading, only one idea remains in the end.

There are more issues to be explored. In multiple rumor spreading the consideration of other types of graphs may lead to important practical results. So far we have avoided the impact of the position of nodes, their study could optimize the spreading by getting through strategical nodes. In the consensus-based approach more work could be realized on the influence of weighted graphs on the convergence of the algorithm. In other words, the updating step could be realized with multiplying the counters by a factor that illustrates the influence of each one of the two communicating nodes.

## VI. ACKNOWLEDGMENT

We would like to thank Vincent Gripon and Michael Rabbat for their supprot and supervision. We would also like to thank Samir Saoudi for his time and consideration. Finally, many thanks to the host laboratory in the electronics departement at Télécom Bretagne.